\documentclass[10pt,conference]{IEEEtran}

\usepackage{cite}

\usepackage[pdftex]{graphicx}
\usepackage{array}
\usepackage{ifthen}
\usepackage{amssymb}
\usepackage{microtype}
\usepackage[normalem]{ulem} 
\usepackage[colorinlistoftodos, textwidth=2.5cm]{todonotes}
\usepackage{multirow}
\usepackage{booktabs}
\usepackage{url}
\usepackage{subcaption}
\usepackage[keeplastbox]{flushend}
\usepackage[T1]{fontenc}

\newboolean{showcomments}
\setboolean{showcomments}{false} 
\ifthenelse{\boolean{showcomments}}
  {
		\newcommand{\nbb}[2]{
		\fcolorbox{black}{yellow}{\bfseries\sffamily\scriptsize#1}
		{\sf$\blacktriangleright$\textcolor{blue}{\textit{#2}}$\blacktriangleleft$}
		}
		
		\newcommand{\remarks}[1]{\color{red}[#1]\color{black}}

		\newcommand{\del}[1]{\textcolor{red}{\sout{#1}}} 
  }
  {
		\newcommand{\nbb}[2]{}
		\newcommand{\remarks}[1]{}

		\newcommand{\del}[1]{} 
  }

\newcolumntype{x}[1]{%
>{\centering\hspace{0pt}}p{#1}}%

\begin{document}
\title{Predicting and Evaluating Software Model Growth\\ in the Automotive Industry}


\author{
\IEEEauthorblockN{Jan Schroeder\IEEEauthorrefmark{1}, Christian Berger\IEEEauthorrefmark{1}, 
Alessia Knauss\IEEEauthorrefmark{1},\\ Harri Preenja\IEEEauthorrefmark{1}, Mohammad Ali
\IEEEauthorrefmark{1},
Miroslaw Staron\IEEEauthorrefmark{1}, Thomas Herpel\IEEEauthorrefmark{7}}
\IEEEauthorblockA{\IEEEauthorrefmark{1}Department of Computer Science and Engineering,
Chalmers and University of Gothenburg, Gothenburg, Sweden,\\
\{jan.schroder; christian.berger; miroslaw.staron; guspreeha; 
mohammad.ali\}@gu.se, alessia.knauss@chalmers.se}
\IEEEauthorblockA{\IEEEauthorrefmark{7}Automotive Safety Technologies GmbH, Ingolstadt, Germany,
thomas.herpel@astech-auto.de}
}

\maketitle

\begin{abstract}
The size of a software artifact influences the software quality and impacts the 
development process. 
In industry, when software size exceeds certain thresholds, memory errors accumulate 
and development tools might not be able to cope anymore, resulting in a lengthy program
start up times, failing builds, or memory problems at unpredictable times.
Thus, foreseeing critical growth in software modules meets a high demand in industrial practice.
Predicting the time when the size grows to the level where maintenance is needed
prevents unexpected efforts and helps to spot problematic artifacts before they 
become critical.

Although the amount of prediction approaches in literature is vast, it is unclear 
how well they fit with prerequisites and expectations from practice.
In this paper, we perform an industrial case study at an automotive manufacturer
to explore applicability and usability of prediction approaches in practice.
In a first step, we collect the most relevant prediction approaches from literature, 
including both, approaches using statistics and machine learning. Furthermore, we 
elicit expectations towards predictions from practitioners using a survey and 
stakeholder workshops. At the same time, we measure software size of 48 software 
artifacts by mining four years of revision history, resulting in 4,547 data points. 
In the last step, we assess the applicability of state-of-the-art prediction
approaches using the collected data by systematically analyzing how well they 
fulfill the practitioners' expectations.

Our main contribution is a comparison of commonly used prediction approaches in a real 
world industrial setting while considering stakeholder expectations. 
We show that the approaches provide significantly different results regarding prediction 
accuracy and that the statistical approaches fit our data best.
\end{abstract}


\IEEEpeerreviewmaketitle

\section{Introduction}
Software development in the automotive industry is facing steadily growing size and
complexity among its artifacts. For example, at Volvo Cars in Sweden, the amount
of software in cars has increased exponentially in the last twenty years: In 2006, 
vehicles contained 10.9MB of code, in 2011 around 117.5MB, and in 2014 already 917MB 
\cite{Hiller16}. Other car companies have seen similar developments and 
according to Wyman~\cite{Wyman15}, the entire automotive sector is facing increasing
technological complexity. Concerning the amount of software and tests being 
introduced with autonomous vehicles, this upwards trend is not going to slow down 
anytime soon.

In practice, software exceeding size limits set by hardware or software requirements
causes long loading, build, and deployment times.
Hence, while 
software grows in size and complexity, situations arise where refactoring and software
maintenance becomes necessary. When these situations occur unexpectedly and immediate 
maintenance become inevitable related delays slow down or stop whole development cycles
and result in increased development costs.
Reliable predictions of software status and quality can prevent such
issues by foreseeing problematic growth in software. This can improve 
release planning processes and provide stakeholders with additional information on 
the evolution of their software.
The need for such predictions at a collaborating automotive manufacturer led us to 
investigate this topic, particularly predicting the size of model-based software,
hereafter software models.


Predicting software model size can be done by assessing growth information 
of past software development. By mining past software revisions and measuring 
the size of the software model time series data is created, which shows the models'
quality development throughout the whole software life cycle. Many approaches for 
predicting time series data exist. Already in 1970, for example, Box and Jenkins~\cite{Box70} 
presented approaches to analyze and predict time series that are used by
stock market or meteorology. 
More recently, machine learning approaches found their way to predict time 
series data as well \cite{Fu11}. In some application domains they have been found 
to outperform classical approaches regarding prediction accuracy \cite{Malhotra15}. 

However, it is still 
unclear how well such approaches perform in a real world, industrial setting where
additional factors affecting their practicability need to be considered like the 
time to train the predictors or their maintainability. Empirical evidence is needed 
to show which approaches are applicable to data that is gathered from realistic 
scenarios in practice.

\subsection{Research Goal and Research Questions}
The goal of this study is to investigate to what extent existing approaches 
are applicable to predict model growth in practice. The following three research 
questions contribute to the goal of our study:
\begin{enumerate}
	\item [RQ1] What are the most important prediction approaches for time series 
        data that are mentioned in the literature?    
    \item [RQ2] Which expectations do practitioners have on the prediction of 
    		software size in practice?
    \item [RQ3] How do the prediction approaches from RQ1 perform 
    		regarding the criteria elicited in RQ2?
\end{enumerate}


\begin{figure}[tb]
  \centering
    \includegraphics[width=\columnwidth]{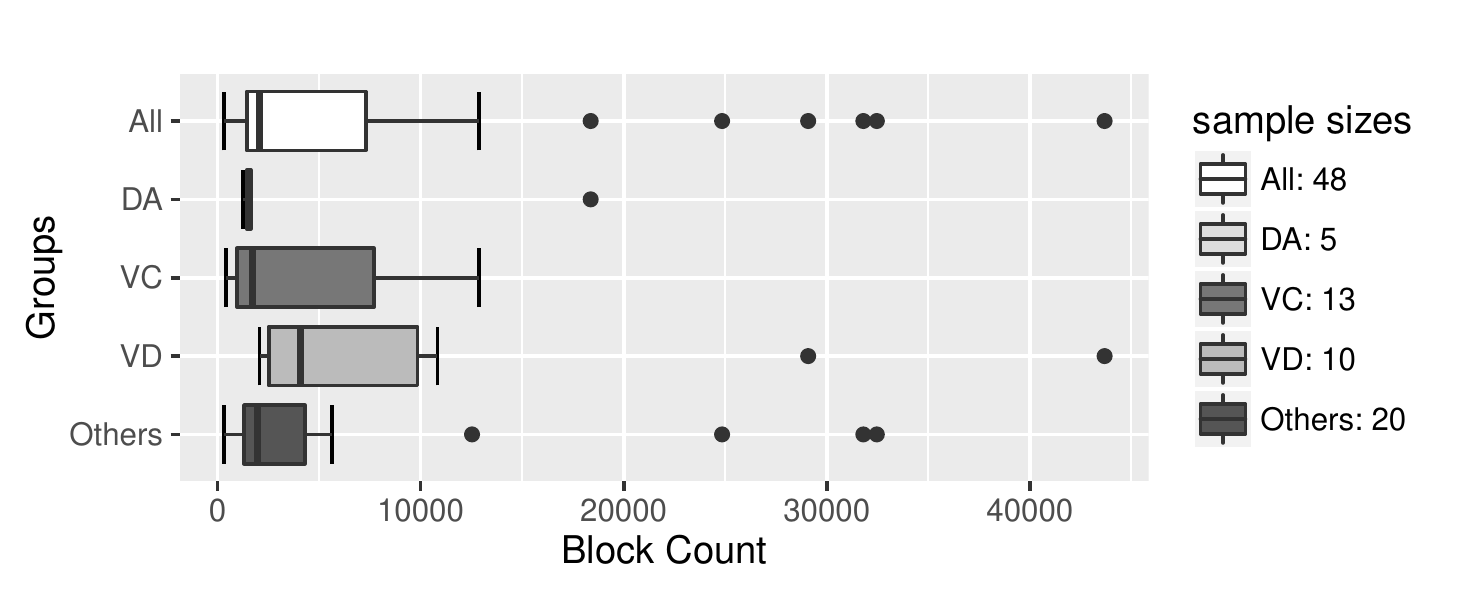}
    \caption{Model size}
    \label{fig:modelOverview}
  \caption{Overview over amount and size (number of blocks) of the models used in 
  this study. 
  }
\end{figure}

\subsection{Contributions}
First, we present five different prediction approaches elicited from
literature and previous studies. They are artificial neural networks (ANN), support 
vector regression (SVR),
long short-term memory (LST), autoregressive integrated moving averages (ARIMA), and 
Holt's linear trend method (HOLT). Second, we list the expectations of ten stakeholders 
collected in questionnaires and workshops. According to the stakeholders in the 
automotive domain, predictions of software growth should be accurate for about one month
period. Finally, this paper contributes by comparing the performance of the five
prediction approaches in an industrial context in the domain of automotive software 
engineering. Our comparison includes traditional statistical approaches (HOLT and ARIMA) 
next to modern machine learning approaches (ANN, SVR, and LSTM) to provide empirical 
evidence about their performance regarding prediction accuracy.
We find that the results received from applying the approaches in practice differ 
significantly from each other. We show that the statistical approaches outperform
machine learning in our context.

\subsection{Paper Structure}
The rest of the paper is structured as follows: We provide background knowledge in 
Section \ref{sec:background} and cover related work in Section \ref{sec:relatedWork}. 
In Section \ref{sec:methodology} we outline our systematic approach to address the 
research questions. Following the methodology, we present, analyze, and 
discuss our results in Section \ref{sec:results}. Section \ref{sec:conclusion} concludes the 
research and presents future work.


\section{Background} \label{sec:background}
\subsection{Context}
This case study is conducted at a testing department of a German premium automobile
manufacturer. The company produces approximately two million vehicles per year. 
In the automotive domain, multiple software projects are combined to 
create the software of a vehicle. Resulting artifacts from these projects are usually
electronic control units (ECUs) to be installed into a vehicle. Simulations of
ECUs are used during testing to replace incomplete real ECUs. This enables the
emulation of a real car environment for ECUs under test. The simulations usually 
run on several real-time computers.

All simulation models needed to simulate a complete vehicle were made accessible to 
the researchers for analysis. The models are realized with Matlab/Simulink. Detailed
information about the simulation models cannot be disclosed, but in order to 
understand the distribution of size and attributes among them, an overview is provided 
in Figure \ref{fig:modelOverview}.
The figure shows the different groups of simulation models 
driver assistance (DA), vehicle control (VC), vehicle dynamics (VD), 
and others. The graphs depict the current model sizes within the groups and the 
whole data set, calculated by counting all blocks in the models.
The sizes of the majority of the models range between 331 and 18,376 blocks, 
with seven positive outliers. The strongest outlier with 107,857 blocks 
within the others-group is not shown in the figure for visibility reasons.

\subsection{Measurements}\label{sec:background:measurements}
Recently, Gil and Lalouche~\cite{Gil17} showed that the measure of size can predict
the validity of any of the 26 metrics they used for comparison. Also, they say
that the higher the correlation with size, the higher the ability of the metric
to estimate external features of the software artifact.
Similarly, our previous studies showed that size metrics are predictors for 
maintainability and software complexity when measuring simulation models
in the automotive industry \cite{Schroeder15}.
They outperformed cohesion and coupling measurements 
in their ability to assess model complexity and maintainability, accordingly.

The increasing size and complexity of software projects can cause severe
problems especially in systems that are limited by the underlying hardware,
for example, in embedded systems. Furthermore, at the case company, lack of 
maintainability among the software models causes significant delays in the 
development process and in the time it takes to introduce new engineers to a project.
In this study, we assess these features by measuring model size in form of lines 
of code (LOC) and block count (BC). 
Both measurements can be considered as static code analysis as they do not
require the models to run for providing results. Hence, they work even if syntax
errors exist in the model. This keeps data preprocessing to a minimum and avoids
unnecessary transformations or interpolation of missing values, which could
skew the data unintentionally.

To calculate the LOC metric in this study, the .mdl files of the Simulink models are 
assessed by counting each line in the XML-like representation of the respective
model. They do not contain comments or similar non-code related entities. The BC
metric counts blocks in a model using the function \emph{sldiagnostics}, which is
built-in in the Matlab/Simulink environment.
The function considers all blocks in the model, even masked blocks on the lowest
layers. In a previous study, we show that both metrics correlate weakly to moderately,
depending on the model under investigation. 
The correlation is not surprising as they both count model size attributes. 
As they are not completely similar, we expect that using two different size metrics in 
this study will provide an additional means for ensuring the validity of the results.

\subsection{Time Series Prediction and Evaluation}

Prediction can be categorized into classification and regression. In classification,
the goal is to assign and learn classes to a set of input values and predict these
classes for each new input value. Regression, on the other hand, aims for learning the
values of some input data and predicting a new value for new or unknown input data. An example 
for classification is predicting nominal categories like \emph{(requirement, feature, bug)} 
for a set of natural language-based issue tracker data. A set of issues would be learned by the
algorithm and a new issue would be predicted to be in one of the three categories.
Predicting regressions usually regards an interval or ratio scale like the 
development of sales over time. Learning sales data of the past enables a prediction
of the sales in the future. 
All approaches presented in this paper are applied to time series data. Time series 
are sequences of observations collected over time, usually in equidistant time 
intervals. In this study, we aim for predicting future values of time series, 
based on previous values of the same time series. Therefore, we use 
regression-based approaches.

Respectively, the approaches for evaluating the performance of predictions differ 
in classification and regression problems.
Measures like F1-measure, Precision, Recall, etc.~work well in classification. 
Contrary to classification problems, the performance of regressions problems
is assessed by how close they approximate a real value. Hence, the data is usually
split into one learning and one test set. Predictions are then made based on the
learning set and compared with the test set. An error measurement is applied
to evaluate the accuracy of the predictions.

\section{Related Work}\label{sec:relatedWork}
In this study, we combine research from the fields of statistical and machine
learning predictions, mining software repositories, and software measurement.

Many existing studies assess different prediction approaches.
Malhotra~\cite{Malhotra15} studied 64 publications regarding the application of machine 
learning techniques for software fault prediction. Malhotra found that 19 out of 
64 studies involved a comparison element between statistical and machine learning
methods. The results demonstrate that the machine learning approaches mostly
outperform statistical linear approaches. The author identified three frequently 
used machine learning approaches for software fault prediction: 1) decision trees, 
2) neural networks, and 3) support vector machines. 
Fu~\cite{Fu11} performed a literature review on prediction approaches as well.
The author provides a comprehensive overview of existing techniques and classifies
them according to their application. There is still a lack of concrete, 
in-depth evaluations of the applicability of the approaches in real-world cases.
Lastly, Mart{\'\i}nez-{\'A}lvarez et~al.~\cite{Martinez15} also review recent work 
on time series prediction. They split results in linear statistical and non-linear
machine learning approaches. We follow their notation in this study. They list 
multiple prediction approaches and error measurements currently used in literature.
Their study, however, is specifically designed for electricity-related time series.
Accordingly, prediction approaches are broadly addressed in literature. Therefore, 
in our study, we investigate the existing work in a literature review to find the
approaches fitting our problem domain.

Size measurement is also well studied. Research investigating software size has shown
that size can be used to assess productivity \cite{Arnold98} and defects 
\cite{Koru05} in practice. Schroeder et al.\cite{Schroeder15} have 
shown that simple metrics like lines of code are well suited to assess software 
complexity and maintainability in a similar environment. Hence, in this study,
we consider size metrics as a powerful and established metric.

Regarding the field of mining software repositories, many studies focus on predicting 
defects like Zimmermann~\cite{Zimmermann09}. Other studies have used software 
repositories to investigate refactoring practices (cf.~\cite{Murphy-Hill12}).
However, to the best knowledge of the authors, no studies have been reported
so far combining aspects of software repository mining and 
measurements with prediction approaches including machine learning approaches
that are evaluated in a realistic industrial context.


\section{Methodology}\label{sec:methodology}
\begin{figure}[!t]
	\centering
	\includegraphics[width=0.49\textwidth]{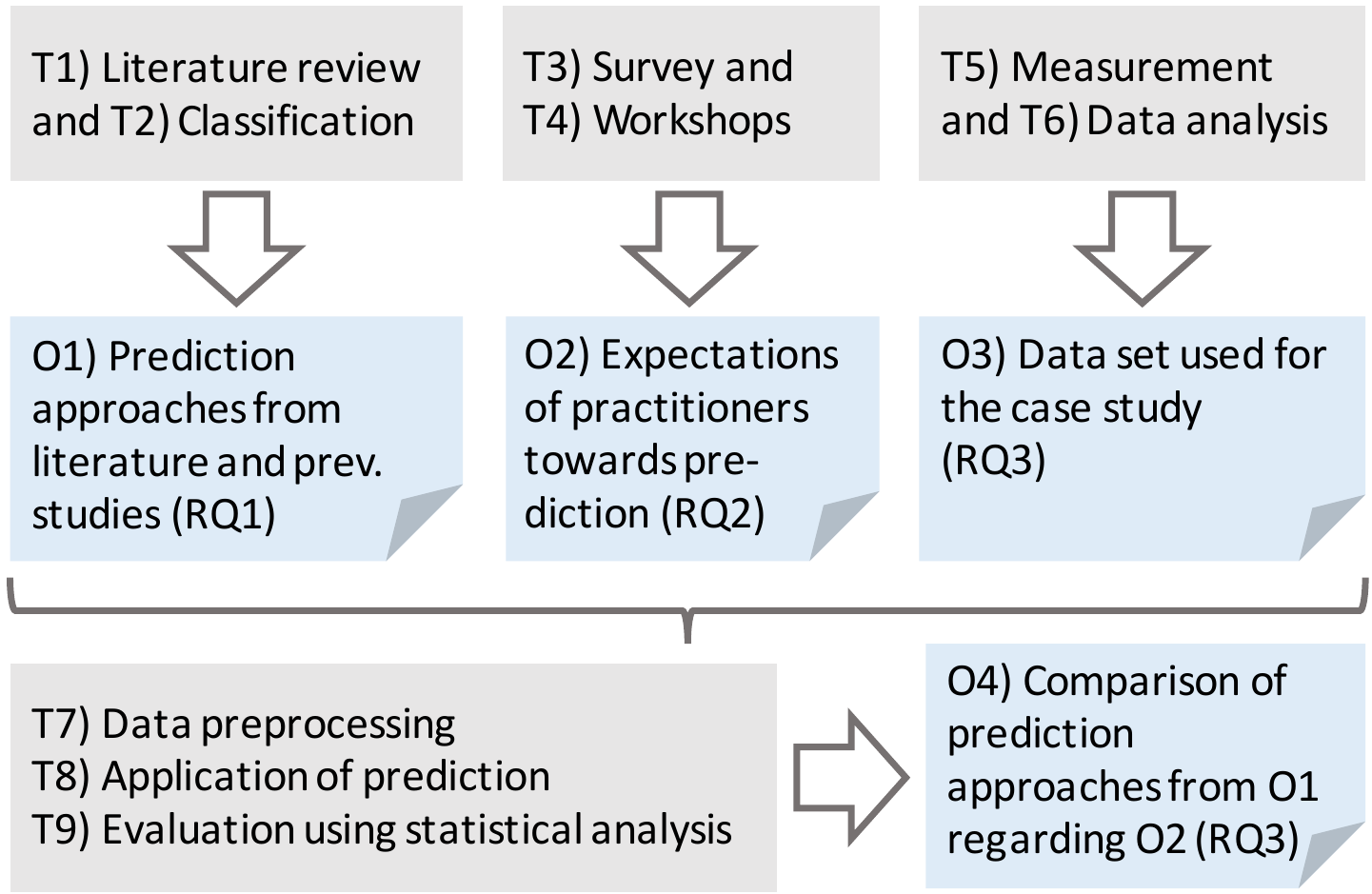}
	\caption{The study design consists of four main tasks. Tasks are presented 
	in gray squared boxes. They are attributed to our research questions. 
	The outcomes are depicted in blue boxes with bent corners.}
	\label{fig:design}
\end{figure}

In this paper, we perform an industrial case study, in which we observe and investigate 
a specific case in a real world context \cite{Runeson12} while avoiding 
interventions of researchers with the case~\cite{Zelkowitz98}. We follow
Runeson and H\"{o}st's guidelines on designing, planning, conducting, and reporting 
the case study \cite{RH08}. Our study comprises quantitative and qualitative methods
and consists of multiple tasks as outlined in Figure \ref{fig:design}.
In the tasks T1-T4, we answer RQ1 and RQ2 using a literature review, a survey,
and workshops. Together with the measurements in task T5, they build the basis 
for the final evaluation of the applicability of prediction approaches in 
industry (RQ3). The tasks are outlined in greater detail in the following sections.

\subsection{Case and Subject Selection}
The artifacts being assessed in this study are Simulink models simulating electronic
control unit (ECU) functionality. All 70 models available at the department are 
considered for the data collection. Most models are still frequently updated with
functionality or quality improvements; other models are only maintained to keep them 
usable in combination with the rest; a third group contains legacy
models which are not used anymore, and the last group is formed by newly created models
with little past data. Legacy models could skew the results as they might not represent
the current development practices. Models that are too new do not provide sufficient
data for our analysis. Hence, as previously shown in Figure \ref{fig:modelOverview},
for this study only the first two groups are considered, and 22 models are removed
from the data set. The remaining 48 models provide a representative view on the 
development conducted for integration testing at the case company. For the 48 models 
4,547 revisions are assessed. This includes only revisions where one of the models' 
code was changed.

Six engineers participated in the industry workshops and ten in the survey. The 
first six engineers have 3.8 years experience on average (1-7 years) and the ten
engineers 4.6 years (1-10 years). The engineers have the roles developer, tester, 
or team leader. All existing roles present at the department are considered for 
the survey. All engineers are working on the shared set of models 
while having different development foci including driver assistance, vehicle 
dynamics, and general vehicle control. All lead developers for the previously 
mentioned function groups are interviewed, as well as the respective team leader.

\subsection{Data Collection Procedure}\label{sec:dataProc}
Our study comprises several tasks involving data collection. They are outlined 
in detail below.

\subsubsection*{T1) Literature Review and T2) Classification}
We conduct a literature review to investigate the most important prediction
approaches. We examine existing literature that focuses on algorithms and
approaches used in the context of predicting time series
data, as well as their implementation in software development. Hence,
the search string is constructed by combining synonyms of the three keywords
\emph{prediction}, \emph{time series}, and \emph{data mining}. We focus on 
research applying prediction approaches. Optimization or in-depth performance
analysis of existing approaches is out of scope of the review. The literature 
studies of Fu~\cite{Fu11}, Malhotra~\cite{Malhotra15}, and 
Mart{\'\i}nez-{\'A}lvarez~et~al.~\cite{Martinez15} are used as main source for
approaches, as they provide existing investigations and comparison.

In the review, we identify the methods and details used for predicting time series.
Based on this information, we identify the most common approaches applied to data 
similar to ours. Further details on methodology and results
of the literature review are published separately.


\subsubsection*{T3) Survey and T4) Workshops}
The goal of the survey and workshops in this study is to complement the
quantitative analysis with practical information from practitioners and to 
provide a qualitative view on software size prediction. The survey in particular
has the purpose to collect the practitioners' expectations on predictions. After
introducing the topic 
the following questions were asked. The possible answers are depicted below 
the questions. The intervals for the answers are based on knowledge received 
during the literature work and gathered at the case company.
\begin{enumerate}
	\item How long can a prediction take, at most?\\
	($<1m$, $<1h$, $<12h$, $\leq 24h$, $>24h$)
	\item How accurate should a prediction be?\\
	(deviation from true values: $1\%$, $3\%$, $5\%$, $10\%$, $\_\_\%$)
	\item How far ahead should the prediction be accurate?\\
	(in days: $1$, $2$, $4$, $7$, $14$, $\_\_$)
    \item How much maintenance effort is acceptable to keep predictions
    running continuously?\\
    (freely specifiable in man-hours per day or week)
    \item What are additional important properties of a prediction?
    (free text)
    \item Rank the prediction properties by their importance:\\
	(\emph{maintenance}, \emph{run time}, \emph{long-term accuracy}, 
    \emph{short-term accuracy},
    and \emph{additional properties} mentioned before)
\end{enumerate}
The questions aim to assess the importance of prediction properties, but also on 
possible quantification of these properties by asking for thresholds. Quantification
is achieved by assigning values to the ranks assigned in the last question, ranging
from one to the number of properties discovered. Those values are counted 
and compared to determine the importance of the properties.

The industry workshops serve the two purposes of discussion and verification of
the gathered results. Findings extracted from the anonymous survey are discussed and
evaluated. The workshops are conducted without a prescribed structure to allow for
discussions about the intermediate results and to find additional input like prediction
properties of interest. This provides input for RQ2 as well as it validates the
applicability of prediction approaches in practice.

\subsubsection*{T5) Measurements and T6) Data Analysis} \label{sec:method:measurements}
\begin{figure}[!t]
	\centering
	\includegraphics[width=0.47\textwidth]{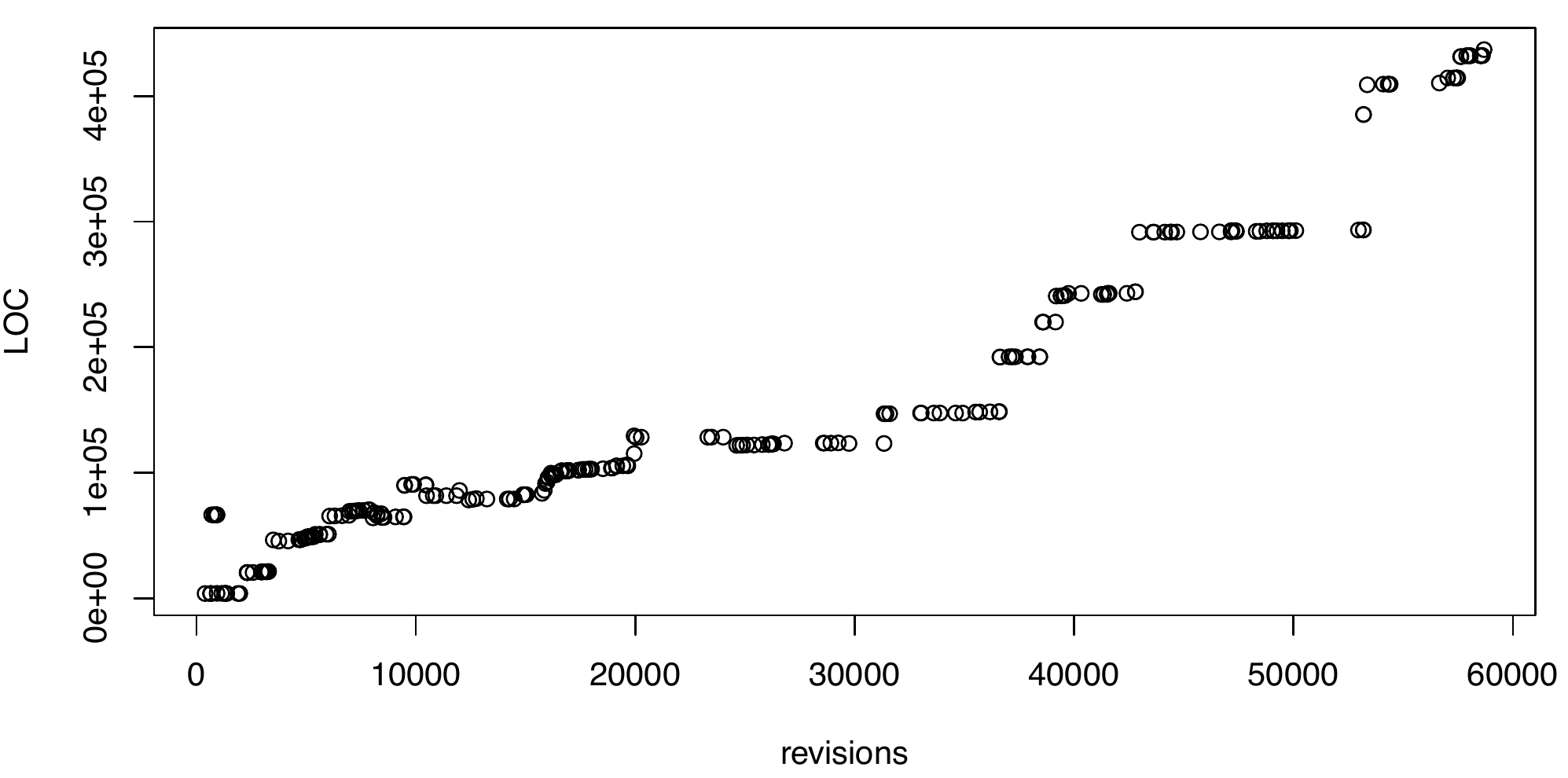}
	\caption{The evolution of model size for one example model 
	measured with the LOC metric.}
	\label{fig:exampleData}
\end{figure}

Having the expectations from practitioners and the background on prediction 
approaches from literature, the data set for the predictions under investigation 
has to be created. The two size metrics described in Section 
\ref{sec:background:measurements} are applied to all 4,547 revisions of the 48 
software models. 
This results in two measurement values, 
one for LOC and one for BC for each revision and therefore two lists for each model.
The lists depict the development of the size metrics over time. On the resulting 
data set, manual data interpretation and time series analysis are performed to understand 
how the data behaves and to fit appropriate prediction approaches later. Investigations include
analysis of trend, seasonality, and outlying/random data. The analysis of the time series
shows that they are mostly determined by their trend. All models grow in size.
We expected seasonal behavior including, for example, regular increases in size within
release cycles but the data set does not express this types of seasonality.
In Figure \ref{fig:exampleData}, the development of the LOC measure is depicted
for one model. The y-axis contains the measurement values and the x-axis the number
of commits.

\subsubsection*{T7) Data Preprocessing}
\begin{figure}[!t]
	\centering
	\includegraphics[width=0.47\textwidth]{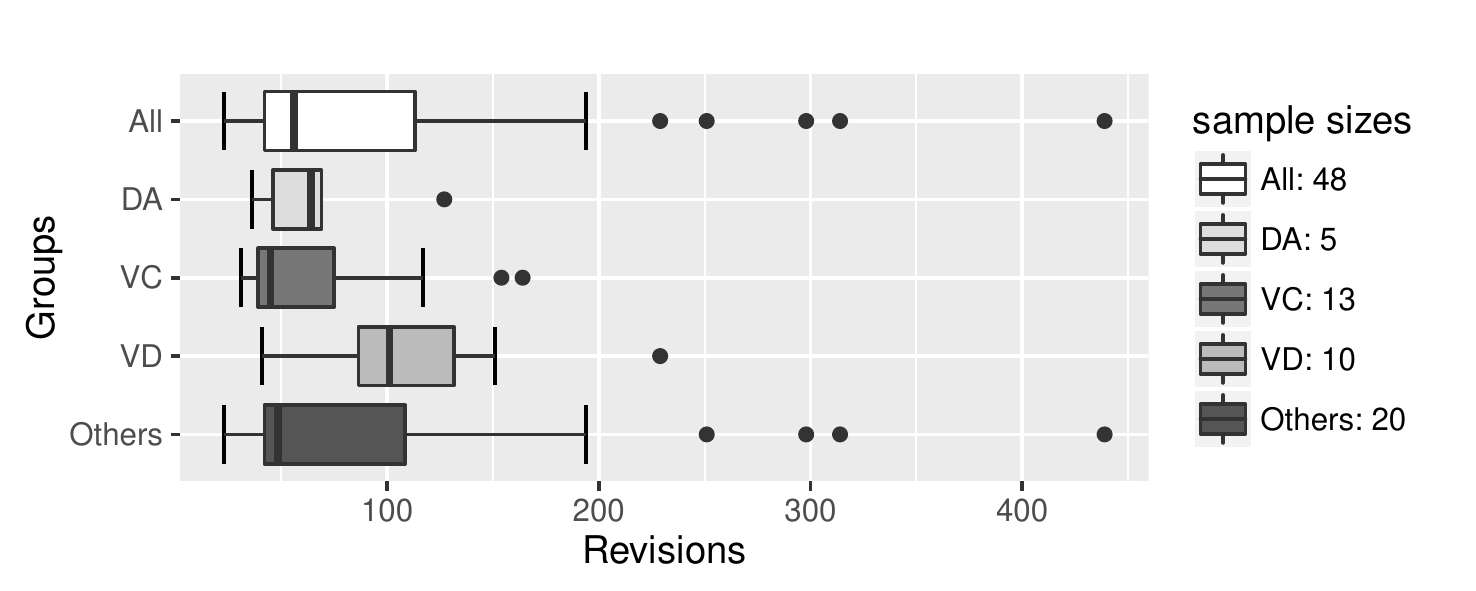}
	\caption{Amounts of model revisions available in this case study, grouped by 
    functionality.}
	\label{fig:revisionDistribution}
\end{figure}

To evaluate the prediction accuracy using ground truth data, the data set,
consisting of measured models throughout revision history, is split up. 
The data is divided into three sets: a training set, a validation set, and a test set. 
The training set is determined by our previous study,
where we collected measurement data between 2013 until 2015. A second data collection was
performed in 2016. This data is split equally into a test and a validation set. 
The resulting data set, which is used for all further analysis in this case study, 
is summarized in Table \ref{tab:revisionData}.
\begin{table}[!t]
\renewcommand{\arraystretch}{1.3}
\caption{Summary of the collected revision data for the 48 models.}
\label{tab:revisionData}
\centering
\begin{tabular}{cccccl}
\hline
 & No. of & Min & Max & Avg. & Period\\
 & revisions & rev. & rev. & & \\
\hline
\hline
All Revisions & 4,547& 20& 439& 94& 1/2013-6/2016 \\
\hline
Learning set&	3,766& 16&	351&	78& 1/2013-12/2015\\
Validation set&	405&	2&	44&	8& 1/2016-3/2016\\
Test set&	376&	2&	44&	7& 4/2016-6/2016\\
\hline
\end{tabular}
\end{table}
Altogether, we collected data from 4,547 revisions in the time between 2013 and 
mid-2016. In this set, we found at least 20 and at most 439 revisions per model. 
Additionally, Figure \ref{fig:revisionDistribution} shows the distribution of all
revisions throughout the groups of functionality.

To perform a time series analysis according to Box and Jenkins~\cite{Box70}
the time intervals between data points have to be equidistant.
However, in the context of revisions and commits, this is not the case, as
commits and consequently measurements are conducted whenever a developer decides to
make changes to the project. There are multiple ways to address this problem.
Eckner~\cite{Eckner12} presents an approach directly applicable to non-equidistant
data and Rehfeld et al.~present a re-sampling approach \cite{Rehfeld11}. Both
are not widely applied. Analyzing time series without constant intervals, 
also called unevenly spaced time series or irregularly sampled time series, 
still requires further research.

Specifically in our case, a more intuitive and straight forward
approach is data interpolation. 
A fixed time interval is chosen, for example, daily intervals. Missing 
values are interpolated. This practice is applied widely but is not without 
critique. A data set can be misinterpreted if it cannot be ensured that in between two 
data points no significant change of values has occurred. This value would
inevitably be missed by interpolation. In our case, it is safe to assume that data 
values actually do not change in between time samples, as we measure every time a 
change to the models was made in form of a commit.
The software does not change in between commits and any missing measurement value can 
be interpolated from the previous data point. This results in time series, which changes
step-wise with each commit while values stay similar in between commits.

We decided to interpolate the data to daily intervals. If multiple commits occurred 
to the same model on a single day we use the latest. According to above descriptions, 
if no commit was made the value for this day is copied from the previous day.
Those daily intervals might introduce bias, as multiple observations during a 
single day are hidden, but it provides a realistic data set for practical observations.

Additional steps like differentiation and normalizing of the time series might be 
necessary. Differentiation removes the trend of a time series and enables
separate investigations of trend and seasonality. Normalization of the data might be
necessary, particularly for the use with neural networks as they are often adjusted 
to work with inputs ranging from 0 to 1.


\subsubsection*{T8) Application of the Predictions}
The selection of prediction approaches is based on the outcome of the 
literature review. The criteria for the selection is how often an approach is
mentioned in literature. The approaches have to be mentioned as being able to handle 
similarly structured data sets as in our case. For our study, both, statistical 
and machine learning approaches are considered.

In addition to the most used approaches, we add two approaches used previously 
on the same data set. In our earlier study, the autoregressive integrated moving 
average (ARIMA) approach was applied. 
As ARIMA is an established 
approach to predict time series, we consider it as an appropriate benchmark. 
Similarly, Holt's linear trend method 
complements
the approaches from literature. Both approaches are explained in more detail together 
with the literature review results in Section \ref{sec:results:RQ1}.


The application of the prediction approaches also has to consider their respective
parameterization. Much time and effort is spent on optimizing parameters of 
predicting approaches to the data at hand. Research is conducted on how to fit
prediction approaches best to a specific data set. Furthermore, for the majority of
the approaches there are parameter estimation algorithms. As it is not the aim of
this study to investigate perfect parameterizations but instead to provide a
practical overview of existing approaches and their applicability in practice,
we decided to use existing parameter estimation algorithms. The libraries providing 
the approaches usually also provide optimization algorithms for parameter estimation.

The selected and configured approaches are then applied to the two lists containing
the measurement data for the LOC and BC metrics created in task T5 for all 4,547 revisions 
of the 48 models. Therefore, we created the predictions for each metric and 
each model, respectively.

\subsection{Analysis Procedure}\label{sec:analysisProc}
In this section, we describe how the data from literature (T1, T2) and the data 
collected in the survey and workshops (T3, T4) is evaluated to answer RQ1 and RQ2, 
respectively. Furthermore, the analysis to determine the performance of the 
prediction approaches is outlined in step T9 to answer RQ3.

\subsubsection*{T1), T2) Classification of Literature and T3), T4) Analysis of 
Survey/Workshop Results}
The approaches found in 
the literature are selected by comparing how often they are mentioned for the use 
on data with a similar structure.
For the assessment of survey results we combine the answers in a table and visualize 
results to make informed decisions on their implications.

\subsubsection*{T9) Evaluation of Prediction Results}
For evaluating regression-based prediction accuracy, we determine how close predicted values
are to the ground truth data. There are multiple possible prediction error measurements
mentioned in literature having different strengths, weaknesses, and biases. Additionally, 
many approaches are very closely related. Based on Adhikari and Agrawal~\cite{Adhikari13} 
and Hyndman and Athanasopoulos~\cite{Hyndman13}, we select the root mean squared 
error (RMSE) as it is widely used and has advantages of making errors comparable
across models. RMSE is calculated by $\sqrt{\sum\frac{(Prediction-Ground Truth)^2}{n}}$ .

\begin{figure}[!t]
	\centering
	\includegraphics[width=0.47\textwidth]{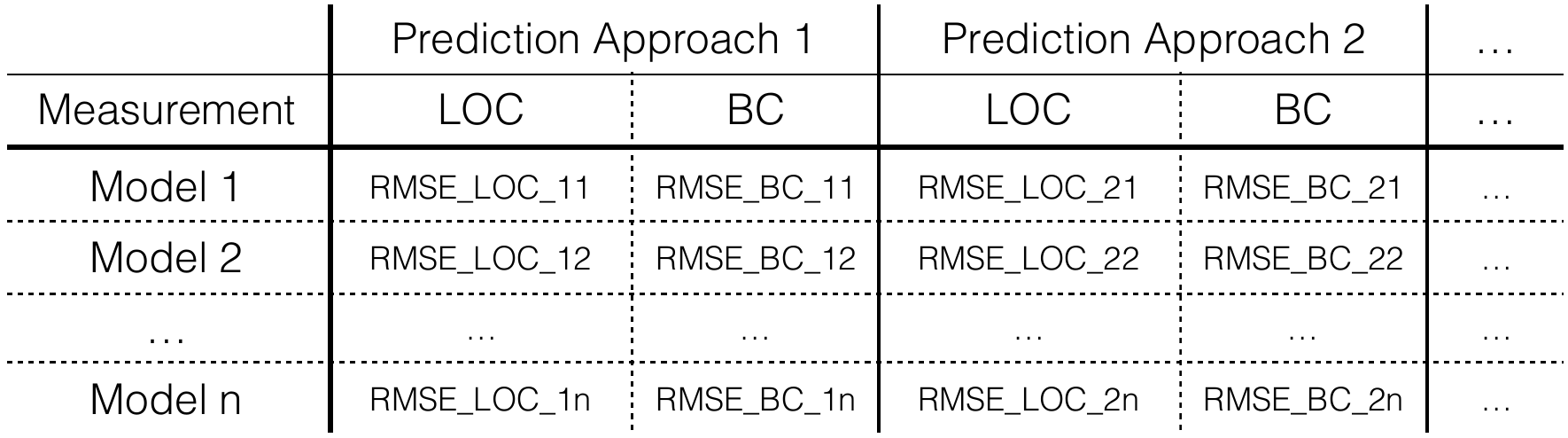}
	\caption{Evaluation of the calculated error measurements. For each approach, the
    measurement results for all models are compared regarding their prediction error.}
	\label{fig:evaluation}
\end{figure}

In order to systematically compare the results, we follow guidelines from 
Basili et al.~\cite{Basili86} and Wohlin et al.~\cite{Wohlin12}. We create a 
hypothesis, determine independent variables, and run statistical tests to find 
statistical significant observations. The comparison of the prediction results is 
visualized in Figure~\ref{fig:evaluation}.
\begin{table*}[tb]
\renewcommand{\arraystretch}{1.3}
\caption{The approaches used in this study and their specifications.}
\label{tab:approaches}
\centering
\begin{tabular}{cccccc}
\hline
Category & Approach & Tool & Library & Parameter estimation algorithm & Parameters\\
\hline
\hline
Statistical, Linear&	HOLT&	R&	forecast&	none &	none\\
Statistical, Linear&	ARIMA&	R&	forecast&	built-in optimization&	AR, I, MA\\
Machine Learning, Non-Linear&	SVR&	Python&	scikit-learn& grid search&
kernel, C, gamma\\
Machine Learning, Non-Linear&	AVNNET&	R&	caret&	caret&	lag, hidden neurons \\
Machine Learning, Non-Linear&	LSTM&	Python&	Keras&	manual and built-in grid search& lag, \# epochs, hidden neurons, optimizer\\
\hline
\end{tabular}
\end{table*}


Using the data presented in this form, the analysis starts with determining the 
distribution of the data to decide if parametric or non-parametric tests should be
used. The next step is comparing the prediction approaches by their prediction 
errors with the goal to determine if there is a statistically significant
difference among the error measurements grouped by the different approaches. Hence,
we define following hypothesis:
\begin{itemize}
	\item[H0] The samples of error measurements for the different approaches 
    originate from identical populations.
    \item[Ha] The samples come from different populations.
\end{itemize}
A one way ANOVA (parametric) or a Kruskal Wallis test (non-parametric) test is used to 
reveal significant differences within the independent groups of prediction errors (RMSE). 
Using these tests, we can show if there is one approach performing significantly better 
or worse than the others regarding prediction errors. This is conducted for both metrics 
individually. Additionally, we also investigate maintenance effort and run time that the
practitioners were asked about in the survey and the workshops.

\subsection{Validity Procedure}
To ensure validity while conducting the study, we focus on employing multiple
combinations of approaches and methods in order to avoid bias. We use five different
prediction approaches with different underlying algorithms, two different size 
metrics, and 48 independent models with and without data interpolation. We are
performing rigorous statistical analysis using established statistical tests. 


\section{Results} \label{sec:results}

\begin{table*}[tb]
\renewcommand{\arraystretch}{1.3}
\caption{The answers collected from the survey, verified in the workshops.}
\label{tab:surveyResults}
\centering
\begin{tabular}{cx{1cm}x{1cm}x{1cm}x{1cm}ccccl}
\hline
 & \multicolumn{4}{c}{Importance Rating} & max time & acceptable & how far to & 
 acceptable maintenance\\
 & accuracy & accuracy & mainte- & run- & to predict & error & predict & \\
 & long & short & nance & time & & \\
\hline
\hline
P1 & 4 & 2 & 3 & 1 & $>$24h & 10\% & 14 days & ``automation, 
big initial effort acceptable''\\
P2 & 2 & 3 & 1 & 4 & $<$1h & 10\% & 21 days & ``less than a man-day per week''\\
P3 & 4 & 3 & 2  & 1 & $<$12h & 10\% & 7 days & ``3h per week''\\
P4 & 4 & 3 & 1  & 2 & $>$24h & 10\% & 30 days & ``automation,
moderate initial effort acceptable''\\
P5 & 4 & 1 & 3  & 2 & $<$1h & 5\% & 90 days & ``1h per week''\\
P6 & 4 & 2 & 3  & 1 & $\leq$24h & 15\% & 28 days & ``2h per 
week''\\
P7 & 3 & 2 & 4 & 1 & $\leq$24h & 5\% & 30 days & ``automation,
only little initial effort acceptable''\\
P8 & 1 & 4 & 3 & 2 & $<$12h & 3\% & 14 days & ``25-30h per week''\\
P9 & 3 & 2 & 4 & 1 & $<$1h & 10\% & 28 days & ``automation''\\
P10 & 1 & 4 & 3 & 2 & $\leq$24h & 5\% & 14 days & ``2-4h per week''\\
\hline
\hline
AVG & 3 & 2.6 & 2.7 & 1.7 & / & 8.3\% & 27.6 days & / \\
\hline
\end{tabular}
\end{table*}

In the results, we consecutively answer the research questions by presenting the 
outcomes of the associated tasks. We present prediction approaches elicited from 
literature and previous studies, followed by the practitioners' expectations 
collected in the survey and workshops. Lastly, we present and analyze the 
prediction results.

\subsection{Predictions approaches Identified in Literature and Previous 
Studies}\label{sec:results:RQ1}
In this section, we answer RQ1, on the most important prediction approaches currently 
used in literature on similar data.
The results of the literature review reveal two major insights. Firstly,
the distinction between linear and non-linear approaches, whereas 
linear approaches are often outperformed by non-linear ones, especially in cases 
when the data exhibits lots of random noise (cf.~\cite{Malhotra15},\cite{Crone06}).

Notably, both linear and non-linear approaches have their strengths and 
weaknesses. The linear approaches exhibit good prediction performance when time
series comprise stationary, non-trending data (cf.~\cite{Garcia04}).
This is because linear approaches predict values based on the previous data in the
time series. To circumvent this weakness, approaches exist to make input data for 
linear approaches stationary. Non-linear approaches such as support 
vector machines (SVM) and artificial neural networks (ANN) have their
strengths in the robustness of the prediction if the data is limited or from 
a short-term period. However, their weakness is a lengthy training process as
mentioned by Sapankevych and Sankar~\cite{Sapankevych09}, Vanajakshi and 
Rilett~\cite{Vanajakshi07}, and Meyfroidt et al.~\cite{Meyfroidt09}.

Secondly, a majority of identified studies highlight the implementation of SVM and 
ANN implemented in a variety of domains. From literature and previous studies, 
we extracted the approaches as listed in Table \ref{tab:approaches}. 
They are briefly outlined in the following paragraphs.

\subsubsection*{Holt's linear trend method (HOLT)}
This approach is serving as a reference for comparison. As shown in Section
\ref{sec:method:measurements} in task T6, the time series are characterized by 
their upwards trend. Smoothing approaches like Holt's are designed for forecasting
trends (cf.~\cite{Hyndman13}). It is provided with R using the ``holt'' function 
contained in the forecast package and does not require configuration. 
The only input required is the time series itself.

\subsubsection*{Autoregressive integrated moving average (ARIMA) model}
ARIMA is an advanced regression approach and commonly used with time series data.
The approach combines an autoregressive function (AR), differentiation (I),
and a moving averages function (MA). These three functions are also the three 
main configuration parameters. ARIMA was used in a previous study with similar 
data to create models for outlier detection.
The R package ``forecast'' provides an
``auto-arima'' function, which compares multiple ARIMA configurations and
selects the configuration according to the model quality. The only input
required is the time series itself.

\subsubsection*{Feed-Forward Artificial Neural Network (ANN)}
Feed-forward artificial neural networks are widely used for time series analysis.
As neural networks are based on learning from past data, it has to be determined
which input should be provided to the network and how. To receive comparable results,
we teach the ANN with the same data as all other approaches: one time series is learned
to create predictions for the same data. For training and 
prediction, we provide the network with a set of past data points, called lagged
data points, instead of feeding one data point at a time. Hence, the network can 
learn to predict a point in the future from multiple past revisions.
We use the ``AVNNET'' implementation provided by the ``caret'' package in R as it
provides automatic parameter estimation functions. The only configuration parameter 
was the size of the input.
As ANNs base upon activation functions, 
typically ranging from 0 to 1, we adjust our data to this range using normalization.

\subsubsection*{Long Short Term Memory (LSTM)}
Long short term memories are recurrent neural networks, which have drawn attention
in the field of forecasting time series in the recent years due to their performance 
(cf.~\cite{Assaad08}, \cite{Ho02}, and \cite{Cheng06}).
An LSTM enhances a plain feed-forward neural network by a memory layer to store
information from a previous learning step and reuse it to influence a current 
learning step.
The ``Keras'' package provides LSTM algorithms for Python. It also provides a grid
search algorithm for parameter estimation. The grid search tries all provided
combinations of configuration parameters and determines the network with the least
error in a given test data. As for the ANN, LSTM is provided with lagged, normalized
input. The following parameters are tried during the grid search: Number of hidden 
neurons, length of lagged input, number of epochs to train, and the optimizer to
be used.

Both neural networks use random weights of the neurons in the beginning of 
learning. Hence, the results created are not deterministic every time the networks 
are trained. To address the non-determinism and receive similar
results for each teaching of the networks, we average the results of multiple
runs following an established method in the field (cf.~\cite{Ripley96}).

\subsubsection*{Support Vector Regression (SVR)}
Support Vector Regression is based on the Support Vector Machine. We implement
it using the ``scikit-learn'' package in Python. The configuration parameters are
the error metric to be used for predictions, the kernel, the penalty parameter 
$C$, and the coefficient for the kernel $gamma$. The scikit-learn package provides 
a grid search algorithm to find the best set of configuration parameters.
\begin{figure*}[tb]
  \centering
  \includegraphics[width=0.975\linewidth]{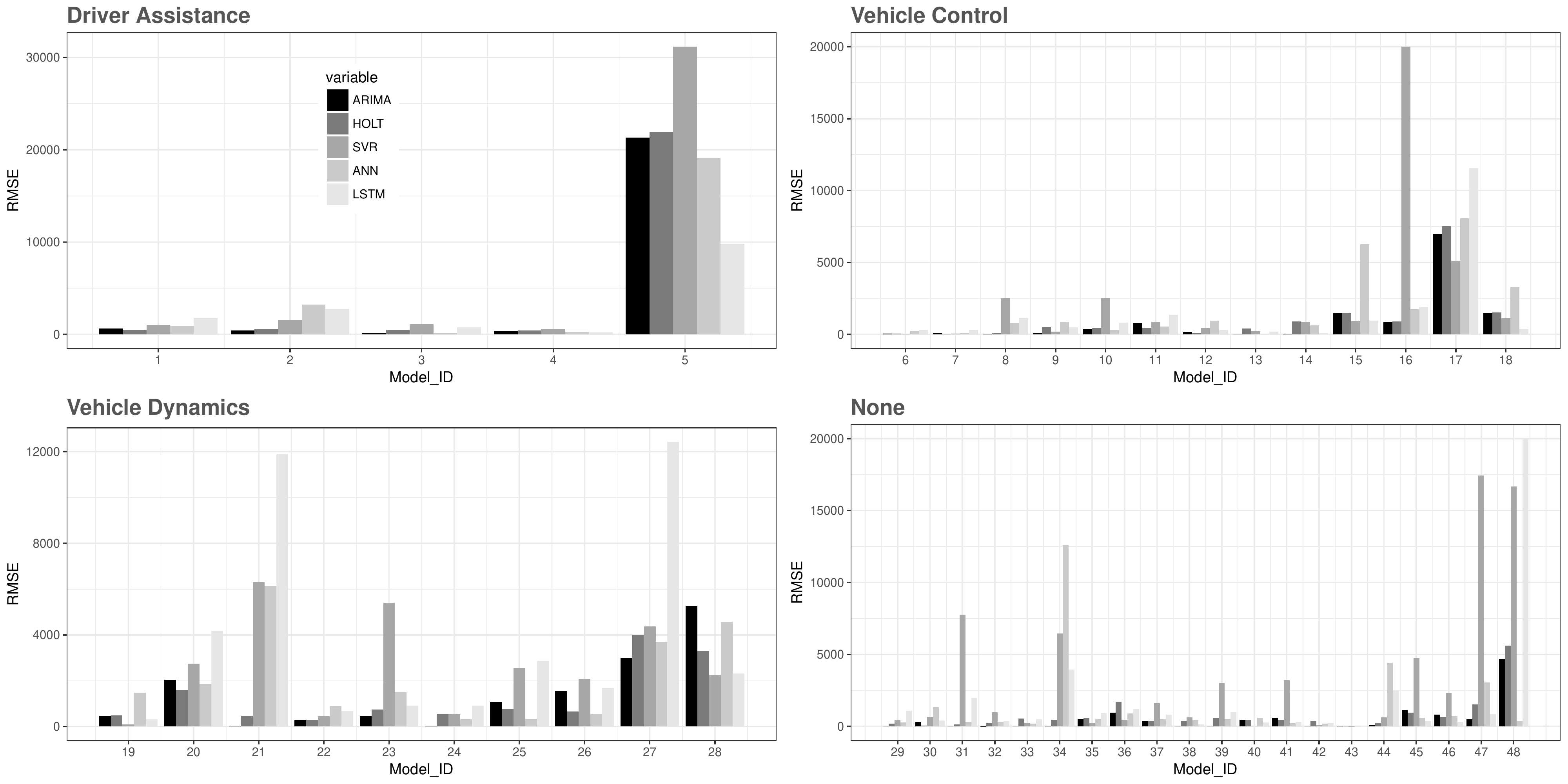}
  \label{fig:modelOverview:size}
  \caption{Overview of the prediction errors (RMSE) by approach  using the LOC metric, 
  arranged in the four groups of functionality.}
  \label{fig:results}
\end{figure*}

\subsection{Expectations of Practitioners towards Predictions} \label{sec:results:RQ2}
In this section we answer RQ2 on the stakeholders' expectations towards predictions.
Findings from the survey are summarized in Table \ref{tab:surveyResults}. 
The table 
shows that predicting over long intervals is most important to the stakeholders. The
respective answer received three points, on average. Still, maintenance and short term predictions 
are almost similarly relevant, with 2.7 and 2.6 points, respectively. Hence, the answers
to the consecutive questions have to clarify the expectations. Interestingly, run time 
received only 1.7 points, on average, and is least important for the stakeholders. The 
related answers for how much time a prediction may take vary strongly. Some engineers
expect a prediction tool which they can have at hand for their work during the day, the 
other group seems to expect predictions to run over night. The answers for accuracy are 
more clear: 
Engineers accept not more than 15\% error within the predictions and 8.3\% on average. 
Both, the results for the error rates and for how far to predict confirm the results of the 
importance rating. It seems that rather high error rates are acceptable as long as predictions 
stay within the boundary, even for long distances. Hence, we interpret that short-term
predictions, for example, until the next day are not sufficient. Engineers expect 
accurate predictions for 27.6 days on average. 

The acceptable maintenance times mentioned 
by the participants highlight the need for automation. Engineers expect the predictions to 
run in an automated way and accept associated initial effort. 
Engineers expect less than three hours of maintenance work on the predictions per week. 
The importance of maintenance time was expected, as man-hours are a valuable resource.
The received results were discussed and validated in consecutive stakeholder workshops.

\subsection{Prediction Results}
In this section we present results from applying the elicited approaches
to the measurements. We answer RQ3 by concluding about the applicability and 
accuracy of the approaches regarding the priorities elicited in the survey and workshops.

\subsubsection{Prediction Accuracy}
The results of the accuracy investigations are summarized in Figure \ref{fig:results}.
The figure shows how the errors computed by RMSE for all five approaches are
distributed among the data. In general, most errors are low while some outlying
spikes are visible. If there are spikes, the approaches mostly increase altogether
like for model 5 and 28. Still, sometimes only the machine learning approaches
are outlying, like in model 16 and 21.

We investigated the prediction accuracy using the engineers' expectations.
According to Table \ref{tab:surveyResults}, on average, practitioners found a deviation 
of 8.3\% from the ground truth acceptable. Hence, we count the number of models where 
predictions, on average, deviate more than 8.3\% from the ground truth.
Table \ref{tab:accuracy10percent} shows the results of this analysis.
\begin{table}[tb]
\renewcommand{\arraystretch}{1.3}
\caption{Results of the accuracy investigations with practitioners' 
expectation of less than 8.3\% error (48 models total).}
\label{tab:accuracy10percent}
\centering
\begin{tabular}{cccccc}
\hline
 & HOLT & ARIMA & SVR & ANN & LSTM\\
\hline
\hline
\# of models above 8.3\% & 7 & 6 & 17 & 11 & 12\\
\hline
\end{tabular}
\end{table}
This data shows the robustness of the approaches by counting the amount of models 
for which the required accuracy could not be achieved. We conclude that the 
approaches HOLT and ARIMA fail in the least cases. SVR predictions deviated 
most from the ground truth data. 
The two statistical approaches seem to outperform the three machine 
learning approaches, considering the stakeholders' expectations. Among the machine 
learning approaches, ANN performs best.

Additionally, we address the hypothesis on a difference between the approaches, 
by performing statistical tests on the error data, as previously shown in 
Figure \ref{fig:evaluation} in Section \ref{sec:analysisProc}. Thereby, we expect
to generate a more reliable and significant evaluation.
The 48 values of RMSE for all five approaches are not normally distributed. 
Hence, we use the Kruskal-Wallis analysis of variance for comparison. This test
compares multiple non-parametric samples to determine if they come from the same population.
The samples in our case are the accuracy results of the predictions, represented as 
RMSE. We test four different data sets: The results for short and long term prediction 
accuracy on both, the LOC and the BC data set, respectively. Whereas for the long term 
the whole set of predictions is used, we consider four revisions into the future 
as short term. Four predictions is the smallest amount of prediction length which 
is achieved by all approaches. Accordingly, for evaluation of short term predictions
the full dataset of 48 models is used. For long term predictions eight data sets are 
excluded. Hence, we have four sets of accuracy results for all five approaches which
can be investigated independently. The result of conducting Kruskal-Wallis tests on 
all four sets are shown in Table \ref{tab:accuracyOverview}.
\begin{table}[tb]
\renewcommand{\arraystretch}{1.3}
\caption{Accuracy comparison using the Kruskal-Wallis test.}
\label{tab:accuracyOverview}
\centering
\begin{tabular}{cccccc}
\hline
&&\multicolumn{2}{c}{short term prediction} & \multicolumn{2}{c}{long term
prediction} \\
&&LOC & BC & LOC & BC \\
\hline
\hline
& HOLT & 110.25 & 112.23 & 114.02 & 113.27\\
Kruskal- & ARIMA & 87.67 & 100.60 & 98.00 & 105.67\\
Wallis & SVR & 144.96 & 151.35 & 137.54 & 146.17\\
Ranks & ANN & 126.38 & 110.54 & 125.02 & 113.75\\
& LSTM & 133.25 & 127.77 & 127.92 & 123.65\\
\hline
\multicolumn{2}{c}{p-values} & 0.001 & 0.004 & 0.059 & 0.043\\
\end{tabular}
\end{table}
The test compares group medians, shown in the center of the table. From comparing
the medians, differences within the error values are visible. Evidence on this 
observation is
provided by the p-values in the lower part of the table. Regarding the short term 
predictions, we can reject the null hypothesis of equal populations. We have evidence 
that at least one approach is significantly different. For long term predictions we 
can reject the null hypothesis in case of the BC metric, both considering a 
significance level $\alpha$ of $0.05$. 

The ranks provided by the test indicate the effect size. 
ARIMA has the lowest medians within RMSE of all approaches, 
while SVR has the highest. By running separate tests in between the groups, 
we found that the predictions performed by SVR are significantly worse than the others.
Between the other groups, no statistically significant difference could be detected.
Lastly, the results suggest that the approaches converge towards long term predictions. 
All approaches decrease in accuracy when predicting long-terms.

\subsubsection{Prediction Maintenance Effort}
The implementations of the approaches in the different languages using different 
libraries strongly depend on expertise with the respective language. Maintenance
effort based on code size or complexity cannot be used as metrics cannot be compared
among different languages. Nevertheless, as we used parameter estimation 
algorithms, which automatically estimate the optimal set of parameters, the 
maintenance effort for future predictions is small for all approaches. As data
changes, the algorithms will find matching sets of parameters. Hence, we
evaluate all approaches to performing equally well.

\subsubsection{Prediction Run Time}
Run time depends on the implementation and the underlying computer system used. 
Still, run time differences can be compared.
Machine learning approaches require a learning period while statistical 
approaches do not require this step. It depends on the accuracy required how
long learning periods have to be and how often they are performed. If run time
is an important aspect, statistical approaches are preferable. In practice
predictions can run during night time. So run time is not an issue in the
context of predicting time series of model growth in the automotive industry.

\subsection{Threats to Validity} \label{sec:threats}
As this case study focuses on one specific case in industry, we assess 
generalizability to other domains as a threat to the validity of the 
results. We mitigated this threat by designing the study in a way to cover multiple
different models, two metrics, and data formats to reduce the chance 
that results are just obtained by chance in the context. We also provide detailed 
descriptions of our data set as well as the approaches and tools used. We expect that 
replications with similar preconditions produce similar results, even in different
domains.

When using machine learning and particularly deep learning approaches, there is always 
the possibility of further optimization. While the results obtained from these
approaches might be improved further, we mitigated this threat to construct
validity by using automatic parameter estimators to ensure a fair treatment
of all approaches.

Furthermore, the approaches might perform differently using other programming
languages or libraries. Even though the approaches should be implemented as
specified by formulas in publications, it could happen that the same approach
is implemented differently in different libraries/tools. Hence, there is a
threat that results can differ when replicating the research with different
libraries or tools. We mitigated this threat by implementing our approaches
with widely used libraries and tools.

Due to the small sample size in the survey, it does only represent a
fraction of the whole industry. We tried to mitigate this threat by considering
practitioners in all different existing roles at the department. Additionally,
the lead developers of all function groups were surveyed. Still, in other domains,
the expectations towards predictions might differ.


\section{Conclusion and Future Work} \label{sec:conclusion}
In this study we compare five prediction approaches, HOLT, ARIMA, SVR, ANN, and LSTM,
which were elicited from literature or used previously on the same data set.
Our results show that all five are applicable to predict time series of software
size measurements performed on simulation models in the automotive industry.

We identify differences regarding their performance, particularly with respect to 
prediction accuracy, which is assessed by calculating the differences between 
predictions and ground truth data. We find that SVR performs significantly worse than 
the other four approaches, in three out of four data sets. 
The data also indicates, that the linear, statistical approaches outperform the
non-linear machine learning approaches regarding accuracy, using our data set. 
This might be due to the comparably small amount of training data available 
(between 6 and 351 data points, 73 on average) or the shape of the data 
that is mostly steadily increasing. For cases with similar data in related application areas 
we recommend linear approaches like HOLT or ARIMA, or feed-forward 
neural networks like ANN.

Additionally, we also contribute by reporting on practitioners expectations towards 
model growth predictions, collected in a survey with ten local developers, testers, 
and team leaders. We find that they expect predictions
to be accurate on long term (about one month) and that short term predictions (about
4 days) of model growth is less important. We also conclude, that predictions are
expected to run automated with low amount of maintenance. Run-times of predictions
are not an issue for most practitioners.



Future investigations are intended to reveal clusters of time series
favoring specific prediction approaches. We might be able to extract general attributes
of data sets that favor specific approaches, like the size of the data set or the 
distance between data points. Accordingly, we would be able to suggest prediction 
approaches matching specific attributes of a dataset.





\bibliographystyle{IEEEtran}
\bibliography{library}

\begin{thebibliography}{10}
\providecommand{\url}[1]{#1}
\csname url@samestyle\endcsname
\providecommand{\newblock}{\relax}
\providecommand{\bibinfo}[2]{#2}
\providecommand{\BIBentrySTDinterwordspacing}{\spaceskip=0pt\relax}
\providecommand{\BIBentryALTinterwordstretchfactor}{4}
\providecommand{\BIBentryALTinterwordspacing}{\spaceskip=\fontdimen2\font plus
\BIBentryALTinterwordstretchfactor\fontdimen3\font minus
  \fontdimen4\font\relax}
\providecommand{\BIBforeignlanguage}[2]{{%
\expandafter\ifx\csname l@#1\endcsname\relax
\typeout{** WARNING: IEEEtran.bst: No hyphenation pattern has been}%
\typeout{** loaded for the language `#1'. Using the pattern for}%
\typeout{** the default language instead.}%
\else
\language=\csname l@#1\endcsname
\fi
#2}}
\providecommand{\BIBdecl}{\relax}
\BIBdecl

\bibitem{Hiller16}
\BIBentryALTinterwordspacing
M.~Hiller, ``{Thoughts on the Future of the Automotive Electronic
  Architecture},'' Presentations at the FUSE Final Seminar, 2016 (accessed
  October 12, 2016). [Online]. Available:
  \url{http://www.fuse-project.se/Homepage/Download-File/f/874242/h/5cfffa89cdaf3a5a46a600d3420b920b/Martin+Hiller+Thoughts+on+the+Future+of+the+Automotive+Electronic+Architecture+%2B+V1.1}
\BIBentrySTDinterwordspacing

\bibitem{Wyman15}
\BIBentryALTinterwordspacing
``A comprehensive study on innovation in the automotive industry,'' Oliver
  Wyman Group, Tech. Rep., 2015. [Online]. Available:
  \url{http://www.oliverwyman.com/content/dam/oliver-wyman/global/en/2014/dec/CarInnovation2015_eng_final.pdf}
\BIBentrySTDinterwordspacing

\bibitem{Box70}
G.~E.~B. Box and G.~M. Jenkins, \emph{\BIBforeignlanguage{English}{Time Series
  Analysis, Forecasting and Control}}.\hskip 1em plus 0.5em minus 0.4em\relax
  San Francisco: Holden Day, 1970.

\bibitem{Fu11}
T.-C. Fu, ``A review on time series data mining,'' \emph{Engineering
  Applications of Artificial Intelligence}, vol.~24, no.~1, pp. 164 -- 181,
  2011.

\bibitem{Malhotra15}
R.~Malhotra, ``A systematic review of machine learning techniques for software
  fault prediction,'' \emph{Applied Soft Computing}, vol.~27, pp. 504 -- 518,
  2015.

\bibitem{Gil17}
Y.~Gil and G.~Lalouche, ``On the correlation between size and metric
  validity,'' \emph{Empirical Software Engineering}, pp. 1--27, 2017.

\bibitem{Schroeder15}
J.~Schroeder, C.~Berger, T.~Herpel, and M.~Staron, ``Comparing the
  applicability of complexity measurements for {S}imulink models during
  integration testing -- an industrial case study,'' in \emph{Proceedings of
  the Second International Workshop on Software Architecture and Metrics
  (SAM)}, May 2015, pp. 35--40.

\bibitem{Martinez15}
F.~Mart{\'\i}nez-{\'A}lvarez, A.~Troncoso, G.~Asencio-Cort{\'e}s, and J.~C.
  Riquelme, ``A survey on data mining techniques applied to electricity-related
  time series forecasting,'' \emph{Energies}, vol.~8, no.~11, pp.
  13\,162--13\,193, 2015.

\bibitem{Arnold98}
M.~Arnold and P.~Pedross, ``Software size measurement and productivity rating
  in a large-scale software development department,'' in \emph{Software
  Engineering, 1998. Proceedings of the 1998 International Conference on}, Apr
  1998, pp. 490--493.

\bibitem{Koru05}
A.~G. Koru and H.~Liu, ``Building effective defect-prediction models in
  practice,'' \emph{IEEE Software}, vol.~22, no.~6, pp. 23--29, Nov 2005.

\bibitem{Zimmermann09}
T.~Zimmermann, ``Changes and bugs -- mining and predicting development
  activities,'' in \emph{Software Maintenance, 2009. ICSM 2009. IEEE
  International Conference on}, Sept 2009, pp. 443--446.

\bibitem{Murphy-Hill12}
E.~Murphy-Hill, C.~Parnin, and A.~P. Black, ``How we refactor, and how we know
  it,'' \emph{IEEE Transactions on Software Engineering}, vol.~38, no.~1, pp.
  5--18, Jan 2012.

\bibitem{Runeson12}
P.~Runeson, M.~Host, A.~Rainer, and B.~Regnell, \emph{Case study research in
  software engineering: Guidelines and examples}.\hskip 1em plus 0.5em minus
  0.4em\relax John Wiley \& Sons, 2012.

\bibitem{Zelkowitz98}
\BIBentryALTinterwordspacing
M.~V. Zelkowitz and D.~R. Wallace, ``Experimental models for validating
  technology,'' \emph{Computer}, vol.~31, no.~5, pp. 23--31, May 1998.
  [Online]. Available: \url{http://dx.doi.org/10.1109/2.675630}
\BIBentrySTDinterwordspacing

\bibitem{RH08}
P.~Runeson and M.~H\"{o}st, ``{Guidelines for conducting and reporting case
  study research in software engineering},'' \emph{Empirical Software
  Engineering}, vol.~14, no.~2, pp. 131--164, Dec. 2008.

\bibitem{Eckner12}
A.~Eckner, ``A framework for the analysis of unevenly spaced time series
  data,'' \emph{Preprint. Available at: http://www.eckner.
  com/papers/unevenly\_spaced\_time\_series\_analysis}, 2012.

\bibitem{Rehfeld11}
K.~Rehfeld, N.~Marwan, J.~Heitzig, and J.~Kurths, ``Comparison of correlation
  analysis techniques for irregularly sampled time series,'' \emph{Nonlinear
  Processes in Geophysics}, vol.~18, no.~3, pp. 389--404, 2011.

\bibitem{Adhikari13}
R.~Adhikari and R.~K. Agrawal, ``An introductory study on time series modeling
  and forecasting,'' \emph{eprint arXiv:1302.6613}, 2013.

\bibitem{Hyndman13}
R.~J. Hyndman and G.~Athanasopoulos, \emph{Forecasting: principles and
  practice}.\hskip 1em plus 0.5em minus 0.4em\relax OTexts: Melbourne,
  Australia, 2013 (Accessed on February 02, 2017).

\bibitem{Basili86}
V.~R. Basili, R.~W. Selby, and D.~H. Hutchens, ``Experimentation in software
  engineering,'' \emph{IEEE Trans. Softw. Eng.}, vol.~12, no.~7, pp. 733--743,
  Jul. 1986.

\bibitem{Wohlin12}
C.~Wohlin, P.~Runeson, M.~H{\"o}st, M.~C. Ohlsson, B.~Regnell, and
  A.~Wessl{\'e}n, \emph{Experimentation in software engineering}.\hskip 1em
  plus 0.5em minus 0.4em\relax Springer Science \& Business Media, 2012.

\bibitem{Crone06}
S.~F. Crone, S.~Lessmann, and S.~Pietsch, ``Forecasting with computational
  intelligence - an evaluation of support vector regression and artificial
  neural networks for time series prediction,'' in \emph{The 2006 IEEE
  International Joint Conference on Neural Network Proceedings}, 2006, pp.
  3159--3166.

\bibitem{Garcia04}
M.~P. Garcia and D.~S. Kirschen, ``Forecasting system imbalance volumes in
  competitive electricity markets,'' in \emph{Power Systems Conference and
  Exposition, 2004. IEEE PES}, Oct 2004, pp. 1805--1812 vol.3.

\bibitem{Sapankevych09}
N.~I. Sapankevych and R.~Sankar, ``Time series prediction using support vector
  machines: A survey,'' \emph{IEEE Computational Intelligence Magazine},
  vol.~4, no.~2, pp. 24--38, May 2009.

\bibitem{Vanajakshi07}
L.~Vanajakshi and L.~R. Rilett, ``Support vector machine technique for the
  short term prediction of travel time,'' in \emph{2007 IEEE Intelligent
  Vehicles Symposium}, June 2007, pp. 600--605.

\bibitem{Meyfroidt09}
G.~Meyfroidt, F.~G{\"u}iza, J.~Ramon, and M.~Bruynooghe, ``Machine learning
  techniques to examine large patient databases,'' \emph{Best Practice \&
  Research Clinical Anaesthesiology}, vol.~23, no.~1, pp. 127 -- 143, 2009.

\bibitem{Assaad08}
M.~Assaad, R.~Bon{\'e}, and H.~Cardot, ``A new boosting algorithm for improved
  time-series forecasting with recurrent neural networks,'' \emph{Information
  Fusion}, vol.~9, no.~1, pp. 41 -- 55, 2008, special Issue on Applications of
  Ensemble Methods.

\bibitem{Ho02}
S.~Ho, M.~Xie, and T.~Goh, ``A comparative study of neural network and
  box-jenkins {ARIMA} modeling in time series prediction,'' \emph{Computers \&
  Industrial Engineering}, vol.~42, no. 2-4, pp. 371 -- 375, 2002.

\bibitem{Cheng06}
H.~Cheng, P.-N. Tan, J.~Gao, and J.~Scripps, \emph{Multistep-Ahead Time Series
  Prediction}.\hskip 1em plus 0.5em minus 0.4em\relax Berlin, Heidelberg:
  Springer Berlin Heidelberg, 2006, pp. 765--774.

\bibitem{Ripley96}
B.~D. Ripley, \emph{Pattern Recognition and Neural Networks}.\hskip 1em plus
  0.5em minus 0.4em\relax Cambridge University Press, 1996.

\end{thebibliography}

\end{document}